\documentclass[aps,pra,twocolumn,superscriptaddress]{revtex4-2}

\usepackage{slashed}
\usepackage{graphicx}
\usepackage{subfigure}
\usepackage[usenames, dvipsnames]{color}
\usepackage{graphics}
\usepackage{hyperref}
\usepackage{bm}
\usepackage{amsfonts}
\hypersetup{backref,
colorlinks=true,
urlcolor=blue,
linkcolor=blue,
linktoc=page,
citecolor=blue}

\everymath{\displaystyle}  
\begin{document}

\title{Dicke States for Accelerated Two Two-Level Atoms 
}

\author{Muzzamal I. Shaukat}
\affiliation{
Texas A\&M University, College Station, Texas 77843, USA}
\email{muzzamalshaukat@tamu.edu}
\author{Charles A. Wallace}
\affiliation{
Texas A\&M University, College Station, Texas 77843, USA}
\author{Anatoly A. Svidzinsky}
\affiliation{
Texas A\&M University, College Station, Texas 77843, USA}
\author{Marlan O. Scully}
\affiliation{
Texas A\&M University, College Station, Texas 77843, USA}
\affiliation{
Baylor University, Waco, Texas 76798, USA
Princeton University, Princeton, New Jersey 08544, USA}
\affiliation{Princeton University, Princeton, New Jersey 08544, USA}

\begin{abstract}
We explore the formation of Dicke states.
A system consisting of
two two-level atoms located in the right Rindler wedge,  has investigated to determine the conditions under which the superradiant or subradiant state can be formed. 
 The dynamics of 
$N$ two-level atoms 
forming 
symmetric state 
has also been analyzed and showed 
that the probability to excite any one atom of a collection of $N$ atoms 
is related to the probability of exciting a single atom.
We derive the analytical expression for the joint excitation probability which demonstrates the 
the interference effect.
These findings provide new insights into the behavior of quantum systems in non-inertial frames and contribute to the broader understanding of relativistic quantum information theory. 
\end{abstract}

\maketitle

\section{Introduction}
The study of quantum systems in non-inertial frames has garnered significant interest due to its implications for both fundamental physics and potential applications in quantum technologies. In this context, the Unruh effect, which predicts that an observer undergoing uniform acceleration perceives the vacuum state as a thermal bath of particles, has been extensively studied both theoretically and experimentally \cite{Fulling1973, Davies1975, Unruh1976}. Quantum optics has emerged as a novel paradigm to study
acceleration radiation, also known as Unruh radiation, in flat/curved spacetime \cite{Scully2003, Scully2018, Anatoly2018}.
Various techniques found in quantum optics have been used to characterize the Unruh effect
such as the quantum master equation technique for the
analysis of black hole acceleration radiation \cite{Scully2018} or excitation through vacuum fluctuation \cite{Anatoly2019}. The study of wave packets and what happens to fermionic Gaussian states under transformation from Minkowski to Rindler coordinates has investigated \cite{Benedikt2017}. Recently, Anatoly et al. have studied the evolution of accelerated harmonic oscillators and have shown how to transform the entanglement of Rindler photons into oscillators moving in causally disconnected regions \cite{Anatoly2025}.
\\
Two-level atoms, despite being the simplest quantum systems, provide a versatile platform for exploring the interaction between matter and fields in such non-inertial settings \cite{ Scully2019, Anatoly2021}. When two such atoms undergo constant acceleration trajectories, their interaction with a massless scalar field can lead to intriguing quantum phenomena, including the formation of entangled states \cite{Grant2015, Arnab2023}.  Dicke states, which are combinations of atomic states (symmetrized or anti-symmetrized), are of particular interest due to their collective quantum properties and potential applications in quantum information science \cite{Peres2004, Fuentes2005}.
Dicke states \cite{Dicke1954} are well-known in the context of inertial frames, where they are employed to describe superradiant and subradiant behaviors of atomic ensembles.
They additionally appear in single-photon superradiance and subradiance effects as well as the collective Lamb Shift \cite{Scully2009, Scully2015}.
\\
 Although the emergence of collective effects, entanglement generation, and spontaneous excitation have been analyzed for an ensemble of uniformly accelerated two-level atoms \cite{Ritcher2017, Wen2018, Audretsch1994, crispino2008},
the formation and properties of Dicke states in non-inertial frames, specifically within Rindler space, remains an open area of research. The unique properties of Rindler spacetime, such as the horizon and the associated thermal nature of the vacuum, introduce novel features into the formation and dynamics of Dicke states that do not have direct analogs in inertial settings.
\\
This paper focuses on the study of two two-level atoms undergoing uniform acceleration in the right Rindler wedge. By analyzing their interaction with a massless scalar field, we aim to understand 
the role of acceleration in obtaining the Dicke Bases and joint excitation Probability.
Through this work, we seek to contribute to the ongoing efforts to
understand quantum field interactions in non-inertial frames, providing insights into how acceleration influences 
excitation processes to form symmetric state for 
N two-level atoms. 
The Dicke states, which are collective quantum states characterized by symmetric and antisymmetric superpositions of atomic excitations, provide a powerful tool for understanding how acceleration influences quantum entanglement and coherence in non-inertial frames.
Future work may explore these effects for different configurations of atomic systems, interactions with various types of quantum fields, and extensions to more general non-inertial scenarios.
\section{Theoretical Model}
\begin{figure}[h]
\centering \includegraphics[scale=0.30]{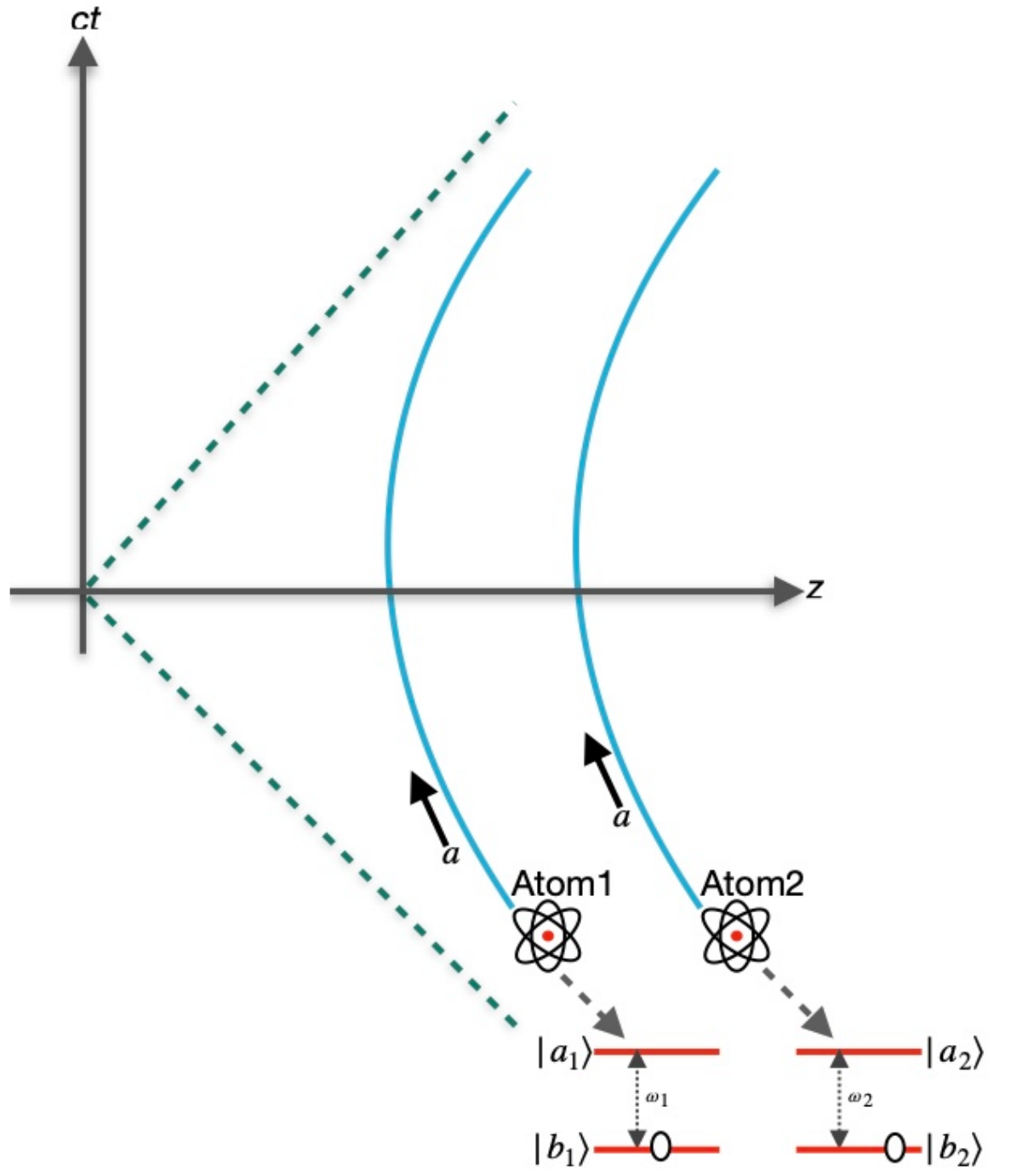}
\caption{ Two atoms are moving with constant acceleration $a$ along z-axis.}
\label{Model}
\end{figure}

We consider two identical, co-accelerating two-level atoms with ground states $\vert b_i\rangle$ ($i=1,2$)
 and excited state $\vert a_i\rangle$
  moving with constant acceleration $a$ 
  in the right Rindler wedge (see Fig. \ref{Model}).  
   For the atoms' proper times $\tau_i \in (-\infty,\infty)$, the Rindler trajectories are given as 
\begin{eqnarray} t_i&=&\frac{c}{a}{\rm sinh}\left( \frac{a \tau_i}{c} \right) \label{Rindler_t}\\  
z_i,
&=&\frac{c^2}{a}{\rm cosh}\left( \frac{a \tau_i}{c} \right)+d_i, \label{Rindler_z}
\end{eqnarray}
where the constant $d_i$ is a coordinate distance added onto a trajectory of either atom.
The interaction Hamiltonian between the atoms and the field can be expressed as
\begin{eqnarray}
H(\tau_i)&=&\hbar \chi \sum_{i=1}^2 \left(\sigma_i e^{-i \omega \tau_i} + \sigma^{\dagger}_i e^{i \omega \tau_i}   \right)
\Phi(t(\tau_i),z(\tau_i))\frac{d\tau_i}{dt}. \nonumber\\
\label{hamiltonian}
\end{eqnarray}
 Here $\chi$ represents the atom-field coupling constant, $\sigma_i$ =  $\vert a_i\rangle \langle b_i \vert$ denotes the atomic energy lowering operator for atom $i$, and its hermitian conjugate is the raising operator of the same, $\omega$ is the transition frequency, the $t$ in the derivative represents Minkowski time as measured by a stationary observer at the origin, and $\Phi(t(\tau_i),z(\tau_i))$ is the field operator evaluated along the trajectory of atom $i$ as prescribed by Eqs (\ref{Rindler_t}) and (\ref{Rindler_z}). The field operator is that of a massless scalar field satisfying the Klein-Gordon equation and is given as
\begin{eqnarray}
\Phi(t,z)&=&
W\left(\hat{a}^R_{\nu} e^{-i(\nu t -k z)}
+ \hat{a}^L_{\nu} e^{-i(\nu t +k z)}+H.C.\right),
\label{field}
\end{eqnarray}
where the R and L refer to the right and left moving modes, respectively, the $z$ and $t$ are the Minkowski coordinates as measured by an observer at the origin, and $W=\sqrt{\hbar /4\pi c \nu}$. We specify the initial state as the ground states for two-level atoms and Minkowski vacuum for the field, i.e., $\Psi_0(t=-\infty)=\vert b_1 b_2 \rangle \vert 0\rangle$.
The evolution operator is $U_i=1-U_I-U_{II}$, where 
\begin{eqnarray}
    U_I&=&\frac{i}{\hbar} \sum_{i=1}^2 \int_{-\infty}^{\infty} d\tau_i H_i(\tau_i(t)), 
  \label{first order}  \\
    U_{II}&=& \frac{1}{\hbar^2}\sum_{i,j}\int_{-\infty}^{\infty}d\tau_i \int_{-\infty}^{\tau}d\tau_j \mathcal{T} H_i(\tau_i(t))H_j(\tau_j(t')), \nonumber\\ \label{second order} 
\end{eqnarray}
where $\mathcal{T}$ is the time ordering operator. Using Hamiltonian Eq. (\ref{hamiltonian}) and  the first-order part of 
 the evolution operator Eq.  (\ref{first order}), we obtain 
 \begin{eqnarray}
\vert \Psi_f^I \rangle &=&U_I\vert \Psi_0\rangle \nonumber\\
&=& i\chi \alpha_+ W\left( e^{ikd_1}|e_1, g_2\rangle+ e^{ ikd_2}  |g_1, e_2\rangle \right)|1^L_{\nu}\rangle \nonumber \\
&+&i \chi \alpha_- W \left( e^{-ikd_1} |e_1, g_2\rangle+ e^{ -ikd_2}  |g_1, e_2\rangle \right)|1^R_{\nu}\rangle,  \nonumber\\ \label{start} \label{Psi_I}
\end{eqnarray}
where $\alpha_{\pm}= \int_{-\infty}^{\infty} d\tau e^{i \omega \tau} e^{\pm i\frac{\nu c}{a}e^{\pm a\tau /c}}$.
The solution of this integral can be determined by employing the following
$\nu t_i\pm kz_i=\pm \frac{c \nu}{a}e^{\pm a \tau_i/c}\pm kd_i 
$, and using change of variables $ x=\nu c/a e^{\pm a \tau_i/c} $ 
to arrive at
\begin{eqnarray}
\alpha_{\pm}&=&   \frac{c}{a} e^{\pm i\theta} 
e^{-\frac{\pi\omega c}{2a}}\Gamma\left(\pm\frac{i\omega c}{a}\right). 
\label{alphas}
\end{eqnarray}
Here $\theta = \omega c/a \ln \big( a/\nu c \big)$ and $\Gamma(\pm ix)=|\Gamma(ix)|e^{\pm i\phi}$ denotes the Gamma function with phase $\phi$. The probability amplitude for a maximally entangled symmetric  ($\vert s \rangle =  \big( \vert e_1 , g_2 \rangle + \vert g_1 , e_2 \rangle \big)/\sqrt{2}$) state with a  right ($-$) or left ($+$) moving photon 
can be determined as
\begin{eqnarray}
\mathcal{A}_{s}^{R(+),L(-)} 
&=&\langle 1_{\nu}^{R(-),L(+)} \vert \langle s \vert \Psi^I_f \rangle
\nonumber\\&=&-i     W  \frac{c \chi}{\sqrt{2}a} e^{\pm i(\theta+kd_1)} 
e^{-\frac{\pi\omega c}{2a}} \nonumber\\
&&(1+e^{\pm ik d})\Gamma\left(\pm \frac{i\omega c}{a}\right).
\end{eqnarray}
Similarly, the probability amplitude for the antisymmetric state and anti-symmetric ($\vert a \rangle =  \big( \vert e_1 , g_2 \rangle - \vert g_1 , e_2 \rangle \big)/\sqrt{2}$) is given by
\begin{eqnarray}
\mathcal{A}_{a}^{R(+),L(-)} 
&=&-i     W  \frac{c \chi}{\sqrt{2}a} e^{\pm i(\theta+kd_1)} 
e^{-\frac{\pi\omega c}{2a}}\nonumber\\
&&(1-e^{\pm ik d})\Gamma\left(\pm \frac{i\omega c}{a}\right).
\end{eqnarray}
where $d = d_2 - d_1$. The probability of determining the symmetric or anti-symmetric state is the sum of the probabilities for the left and right moving modes, i.e.,
\begin{eqnarray}
    \mathcal{P}_s&=&
\frac{8\pi c \chi^2 W^2}{\omega a} \frac{{\rm cos}^2 (\frac{kd}{2})  }{ (e^{2\pi\omega c/a}-1)}. \label{P_s}
 \\
\mathcal{P}_a&=&\frac{8\pi c \chi^2 W^2}{\omega a} \frac{{\rm sin}^2 (\frac{kd}{2})  }{ (e^{2\pi\omega c/a}-1)}. \label{P_a}
\end{eqnarray}
Here we used $|\Gamma(-i x)|^2=\pi/(x {\rm sinh}(\pi x))$. 
It can be observed that the excitation probability depends on the distance between two atoms and Planck factor with atomic transition frequency $\omega$ and Unruh temperature $T_U=\hbar a/2\pi ck_B$. $\mathcal{P}_s$ and $\mathcal{P}_a$ determine the constructive and destructive interference effect encoded in trigonometric terms, which depends on separation
$d$.
Furthermore,
the probability of exactly one atom of the pair being excited and emitting a photon is given by the sum of $\mathcal{P}_s$ and $\mathcal{P}_a$ and is twice the probability of exciting a single atom executing the same trajectory. 
This model can be extended to obtain
the symmetric state for N two-level atoms by considering 
all the atoms  in the ground state initially,  i.e, 
\begin{eqnarray}
    \vert \psi^{IN}_f \rangle &=& -i W\chi \sqrt{N}
   ( \alpha_+ \vert S^+ \rangle \vert 1_{\nu,L} \rangle+\alpha_- \vert S^- \rangle \vert 1_{\nu,R} \rangle), 
\end{eqnarray}
where $\vert S^{\pm}\rangle=\sum_{j=1}^N e^{\pm ikd_j} \vert g_1, g
_2,...,e_j g_{j+1},g_{j+2}...\rangle/\sqrt{N}$.
The probability to excite exactly one atom of a collection of $N$ atoms 
is related to the probability of exciting a single atom executing the same trajectory 
\begin{equation}
    \text{Pr}(\text{1 excited of $N$}) = N \text{Pr}(\text{lone atom excited}).
\end{equation}
\begin{figure}[h]
\centering \includegraphics[scale=0.60]{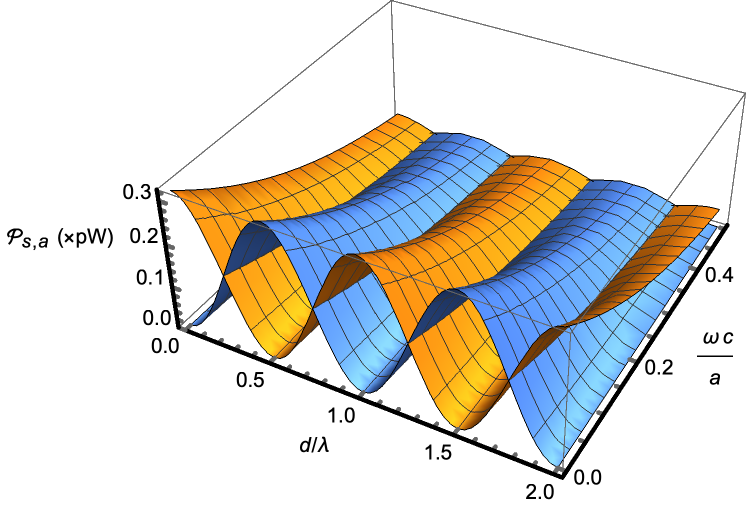}
\caption{Initial maximum probability shows the constructive interference for the symmetric state while the blue curve determines the destructive interference for the anti-symmetric state. The parameters are $k=2\pi/\lambda$, atom-field coulping constant  $\chi=10MHz$, photon frequency $\nu=0.1\omega$ and transition frequency $\omega= 1GHz$. 
}
\label{interference1}
\end{figure}
\\
The probability of excitation of  two atoms with the emission of two photons is due to 
counter-rotating atomic operators $\sigma_i^{\dagger}(\tau) \sigma_j^{\dagger}(\tau')$ $(i\neq j)$ and field operators $
  a_m^{\dagger}(\tau) a_n^{\dagger}(\tau') $ $(m=(R,L)=n)$. Applying these operators on the initial state $\vert \Psi_0\rangle$ determines the probability amplitude for the doubly excited state with the emission of two photons in any direction 
\begin{eqnarray}
  \vert \Psi^{II}_f \rangle &=& U_{II} | \Psi_0 \rangle \nonumber \\
  &=& W^2 \chi^2\Big( 2\sqrt{2} e^{ik(d_1 + d_2)} \beta_{LL} \vert 2^{L}_{\nu} \rangle \nonumber \\
  &+& 2\sqrt{2} e^{-ik(d_1 + d_2)} \beta_{RR} \vert 2_{\nu}^{R} \rangle  \label{Psi_II}\nonumber \\
  &+& \big[ (\beta_{RL}+\beta_{LR}){\rm cos}(kd)
  \big] \vert 1_{\nu}^R 1_{\nu}^L \rangle 
  \Big) \vert e_1 e_2 \rangle ,
\end{eqnarray}
where 
\begin{eqnarray}
\beta_{LL}&=& 
\int_{-\infty}^{\infty} \int_{-\infty}^{\tau} e^{i \frac{\nu c}{a} e^{\frac{a\tau}{c}}} e^{i \frac{\nu c}{a} e^{\frac{a\tau'}{c}}} e^{i \omega_1 \tau} e^{i \omega_2 \tau'} d\tau' d\tau
\nonumber
\label{2leftApp}
\end{eqnarray}
represents the probability amplitude for two left-moving photons. We make the substitutions
\begin{eqnarray}
    x = \frac{\nu c}{a} e^{\frac{a \tau}{c}}, 
  \hspace{1cm} {\rm and}  \hspace{1cm}  y = \frac{\nu c}{a} e^{\frac{a \tau'}{c}}, \label{xy_sub}
\end{eqnarray}
and assumes that the atoms are identical for which we set $\omega_1=\omega_2=\omega$ and 
the acceleration is positive $a_1=a_2=a$. Therefore, the substitutions result in the integral
\begin{equation}
   \beta_{LL}= \frac{c^2}{a^2} e^{-2i\theta}\int_0^{\infty} \int_0^{x} e^{ix} e^{iy} x^{i \frac{\omega c}{a} -1} y^{i \frac{\omega c}{a} -1} dy dx. \label{leftamp}
\end{equation}
Taking advantage of the expansion of lower incomplete Gamma function as
\begin{eqnarray}
\gamma(s,z)&=&\int_0^z e^{-u}u^{s-1}du \nonumber \\
&=&\sum_{k=0}^{\infty}\frac{(-1)^k z^{k+s}}{k! (k+s)},
\end{eqnarray}
and using the exponential Taylor series in the $y$ integral,
\begin{equation}
  \sum_{k=0}^{\infty} \frac{i^k}{k!} \int_0^x y^{i\frac{\omega c}{a} + k -1} dy = \sum_{k=0}^{\infty} \frac{i^k}{k! (k + i\frac{\omega c}{a})} x^{i\frac{\omega c}{a} + k}.
\end{equation} 
Eq. (16) can be simplified to obtain
\begin{eqnarray}
  \beta_{LL}&=&
\frac{c }{i a\omega}  e^{-2i\theta} e^{-\frac{\pi\omega c}{a}}\Gamma\left(\frac{2i\omega c}{ a}\right) \nonumber\\ &&
{_{2}}F_1\left[\frac{2ic\omega}{a},\frac{ic\omega}{a};\frac{ic\omega}{a}+1;-1 \right]. 
  \label{2left}
\end{eqnarray}
Here ${_{2}}F_1$ represents the Hypergeometric Function. 
Using the property ${_{2}}F_1[a,b;1+a-b;-1]=\Gamma(1+a-b)\Gamma(1+a/2)/(\Gamma(1+a/2-b)\Gamma(1+a))$ \cite{Andrews1999} and $\Gamma(1+z)=z\Gamma(z)$, Eq. (\ref{2left}) is given by 
\begin{eqnarray}
\beta_{LL}&=&\frac{\pi c }{\omega a} \frac{e^{-2i(\theta-\phi) }  }{ (e^{2\pi\omega c/a}-1)}.
 \label{2left21}
\end{eqnarray}
Similarly, the probability amplitude for two right-moving photons is given by
\begin{eqnarray}
\beta_{RR}&=&\int_{-\infty}^{\infty} \int_{-\infty}^{\tau} e^{-i \frac{\nu c}{a} e^{-\frac{a\tau}{c}}} e^{-i \frac{\nu c}{a} e^{-\frac{a\tau'}{c}}} e^{i \omega_1 \tau} e^{i \omega_2 \tau'} d\tau' d\tau 
\nonumber\\
&=&\frac{\pi c }{\omega a} \frac{e^{2i(\theta-\phi) }  }{ (e^{2\pi\omega c/a}-1)}.
\label{2right21}
\end{eqnarray}
For one left and one right moving photons, two integrals will appear which will be associated with either applying the left or right creation operator (i.e., $a_R^{\dagger}(\tau) a_L^{\dagger}(\tau')  ,   a_L^{\dagger}(\tau) a_R^{\dagger}(\tau') \hspace{1mm} $). Beginning with the right creation operator applied first and following the procedure outlined for two left-moving photons, we obtain
\begin{eqnarray}
    \beta_{RL} &=& \int_{-\infty}^{\infty} \int_{-\infty}^{\tau} e^{-i \frac{\nu c}{a}e^{-\frac{a\tau}{c}}} e^{i\frac{\nu c}{a} e^{\frac{a \tau'}{c}}} e^{i\omega (\tau + \tau')} d\tau' d\tau, \nonumber \\
    &=& \frac{ce^{-2i \theta}}{i\omega a} e^{-\frac{\pi \omega c}{a}} \Gamma\Big( -i\frac{2\omega c}{a} \Big) H.
\end{eqnarray}
where $H={_{1}}F_2 \left[ -i\frac{\omega c}{a} ; -i\frac{\omega c}{a} + 1 , -2i \frac{\omega  c}{a} ; -\left(\frac{\nu c}{a}\right)^2 \right]$. It is pertinent to mention here that $ \beta_{LR} = \vert \beta_{LR}\vert e^{i\phi_{LR}} = (\beta_{RL})^*$.
Then the probability of finding the atomic system in the doubly excited state is given by 
\begin{eqnarray}
P_{e_1 e_2}
&=& \frac{16 \pi c W^4 \chi^4 }{\omega^3 a} \Big[\frac{\pi \omega c}{a}
\frac{1}{\left(e^{\frac{2\pi \omega  c}{a} }-1\right)^2}
\nonumber \\
&-&   \frac{\cos^2 (kd)\cos^2(\phi_{RL})}{\left(e^{\frac{4\pi \omega c}{a} }-1\right)}
|H|^2\Big].
\end{eqnarray}
It is worth noting that the first term shows that an observer in an accelerated frame detects the quantum vacuum as a thermal bath with a temperature proportional to acceleration $a$ due to the Unruh effect and the second term contains the cosine function with spatial separation $d$ and phase difference $\phi_{RL}$ rendering
the interference effect. Therefore, the final state is given by
\begin{eqnarray}
\vert \Psi_F\rangle
&=& \vert \Psi_0   \rangle + \vert \Psi_f^I \rangle + \vert \Psi_f^{II} \rangle,
\label{final state}
\end{eqnarray}
where $\vert \Psi_f^I \rangle$ and $\vert \Psi_f^{II} \rangle$ are given by Eq.'s (\ref{Psi_I}) and (\ref{Psi_II}), respectively. It is pertinent to mention here that the term $\left(\vert e_1 g_2\rangle + e^{\pm ikd} \vert g_1 e_2\rangle   \right)$ in  $\vert \Psi_f^I \rangle$ can be simplified by considering  $kd=n\pi=n\lambda/2$, which 
creates the entangled symmetric state $\vert s \rangle$ for even values of n and anti-symmetric state $\vert a \rangle$ for odd values of n.
However, the probability of excitation for symmetric and antisymmetric states depending on distance $d$ is described in Eq.'s (\ref{P_s}) and (\ref{P_a}). 
Therefore, Eq. (\ref{final state}) forms a set of Dicke bases, where $\vert g\rangle=\vert b_1 b_2\rangle$ and $\vert e\rangle=\vert a_1 a_2\rangle$, $\vert s \rangle=\left(\vert a_1 b_2\rangle +  \vert b_1 a_2\rangle   \right)/\sqrt{2}$ and $\vert a \rangle=\left(\vert a_1 b_2\rangle -  \vert b_1 a_2\rangle   \right)/\sqrt{2}$. 
This perturbative framework is essential not only for the theoretical modeling of collective quantum phenomena but also for the practical realization of robust and scalable entangled states in quantum technologies.
\begin{figure}[h]
   \centering
    \includegraphics[scale=0.45]{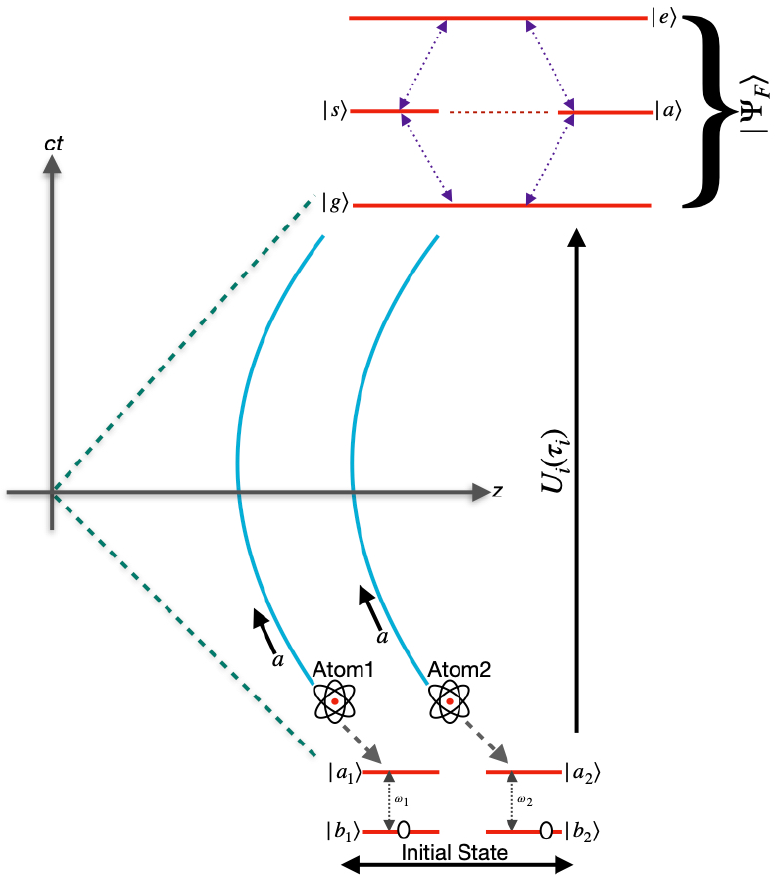}
    \caption{ Dicke States of two two-level atoms.  
    The ground state $\vert b\rangle$ has both atoms in the ground state, initially.
The next state has one atom excited which generates the maximally entangled symmetric $\vert s\rangle$ and anti-symmetric $\vert a\rangle$ states. This process continues until both atoms are excited $\vert e\rangle$.    
    }
\label{Dicke}
\end{figure}
\section{Summary and Discussion}
In summary, we have investigated how to formulate the Dicke basis for a system of two uniformly accelerated two-level atoms in the right Rindler wedge that interact with a quantum field. 
We explored how the Unruh effect, which manifests as a thermal bath experienced by accelerated observers, influences the collective quantum states of the atoms.
 We demonstrated that the symmetric and antisymmetric Dicke states exhibit distinct excitation probabilities proportional to the Planck factor with atomic frequency, Unruh temperature and highlighting the dependence on the separation between atoms to reveal the constructive and destructive interference effect.
  We also determined the symmetric state for $N$-two-level atoms and showed that this probability is $N$ times the probability for a single two-level atom in Rindler space.
We further intend to excite both atoms to 
derive the joint excitation probability and extract the dependence on the separation between atoms to render an interference effect. Future work may explore these effects for different configurations
of atomic systems, interactions with various types of quantum fields, and extensions to more general non-inertial
scenarios.
\section*{Acknowledgements}
This work is supported by the Welch Foundation (A-1261), the DARPA PhENOM program, the Air Force Office of Scientific Research (Award No. FA9550-20-1-0366), and the National Science Foundation (Grant No. PHY-2013771). This work is also supported by the US Department of Energy under award numbers DE-SC-0023103, DE-SC0024882, and FWP-ERW7011.

\end{document}